\documentclass[11pt,american,english]{article}
\usepackage[T1]{fontenc}
\usepackage[latin9]{inputenc}
\usepackage[a4paper]{geometry}
\usepackage{color}
\usepackage{babel}
\usepackage{amsmath}
\usepackage{amssymb}
\usepackage{graphicx}
\usepackage{setspace}
\usepackage[numbers]{natbib}
\usepackage[unicode=true,
 bookmarks=true,bookmarksnumbered=true,bookmarksopen=true,bookmarksopenlevel=3,
 breaklinks=true,pdfborder={0 0 0},backref=page,colorlinks=true]
 {hyperref}
\hypersetup{pdftitle={MIRS_2011},
 pdfauthor={J. P. Champion}}

\makeatletter

\providecommand{\tabularnewline}{\\}

\newcommand{\lyxaddress}[1]{
\par {\raggedright #1
\vspace{1.4em}
\noindent\par}
}

\date{}

\makeatother

\begin{document}

\title{Extension of the \foreignlanguage{american}{MIRS} computer package
for the modeling of molecular spectra : from effective to full \textit{ab
initio} ro-vibrational hamiltonians in irreducible tensor form}

\author{A. V. Nikitin$^{a}$, M. Rey$^{b}$, J. P. \textsc{\emph{Champion}}$^{c}$
and Vl. G. Tyuterev$^{b}$ }

\maketitle

\lyxaddress{\begin{center}
$^{a}$Laboratory of Theoretical Spectroscopy, Institute of Atmospheric
Optics, Russian Academy of Sciences, 634055, Tomsk, Russia.\\
 $^{b}$Laboratoire de Spectroscopie Moléculaire Atmosphérique (UMR
CNRS 6089), Université de Reims, B.P. 1039, 51623, Reims, France.\\
 $^{c}$Laboratoire Interdisciplinaire Carnot de Bourgogne (ICB),
(UMR CNRS 6303) - Université de Bourgogne, 9 Av. A. Savary, BP 47870,
F-21078 DIJON, France.
\par\end{center}}
\begin{abstract}
The MIRS software for the modeling of ro-vibrational spectra of polyatomic
molecules was considerably extended and improved. The original version
(Nikitin, et al. JQSRT, 2003, pp. 239--249) was especially designed
for separate or simultaneous treatments of complex band systems of
polyatomic molecules. It was set up in the frame of effective polyad
models by using algorithms based on advanced group theory algebra
to take full account of symmetry properties. It has been successfully
used for predictions and data fitting (positions and intensities)
of numerous spectra of symmetric and spherical top molecules within
the vibration extrapolation scheme. The new version offers more advanced
possibilities for spectra calculations and modeling by getting rid
of several previous limitations particularly for the size of polyads
and the number of tensors involved. It allows dealing with overlapping
polyads and includes more efficient and faster algorithms for the
calculation of coefficients related to molecular symmetry properties
($6C$, $9C$ and $12C$ symbols for $C_{3v}$, $T_{d}$, and $O_{h}$
point groups) and for better convergence of least-square-fit iterations
as well. The new version is not limited to polyad effective models.
It also allows direct predictions using full \textit{ab initio} ro-vibrational
normal mode hamiltonians converted into the irreducible tensor form.
Illustrative examples on CH$_{3}$D, CH$_{4}$, CH$_{3}$Cl, CH$_{3}$F
and PH$_{3}$ are reported reflecting the present status of data available.
It is written in C++ for standard PC computer operating under Windows.
The full package including on-line documentation and recent data are
freely available at \href{http://www.iao.ru/mirs/mirs.htm}{http://www.iao.ru/mirs/mirs.htm}
or \href{http://xeon.univ-reims.fr/Mirs/}{http://xeon.univ-reims.fr/Mirs/}
or \href{http://icb.u-bourgogne.fr/OMR/SMA/SHTDS/MIRS.html}{http://icb.u-bourgogne.fr/OMR/SMA/SHTDS/MIRS.html}.
\end{abstract}

\newpage{}

\tableofcontents{}

\listoffigures

\listoftables
\newpage{}

\section{Introduction}

Spectroscopic investigations of polyatomic molecules play a role of
primordial importance for atmospheric applications. Effective hamiltonians
in irreducible tensor operator (ITO) form have been successfully used
to model the crucial effects of ro-vibrational perturbations on line
positions and strengths through the analysis of complex interacting
band systems (so-called polyads). An overview can be found in \citep{Champion_1992,Hilico_2001,Rotger_2000a,Nikitin_2003a,Nikitin_2004,Boudon_2004,Nikitin_2005a,Nikitin_2006,Rotger_2002,Nikitin_2002,Albert_2009}
and references therein. This approach proved to be particularly efficient
for a full account of the molecular and space symmetry properties.
Hereafter we shall refer to these models as ITO spectroscopic models. 

Nowadays, modern high-resolution and high-sensitivity techniques as
well as direct spectroscopic observations of the planetary atmospheres
and stars yield a wealth of spectroscopic data involving highly excited
energy levels and relatively weak spectroscopic features. The modeling
of such data using empirically adjusted effective hamiltonians becomes
more and more complicated for several reasons. Most vibration-rotation
bands show irregular perturbations due to accidental resonances. The
importance of such effects in high-resolution spectroscopy has been
shown in works by J.-M. Flaud, C. Camy-Peyret and co-workers \citep{FLAUD_1975,Camy-Peyret_1985}
who were among the first researchers developping effective model hamiltonians
and dipole transition moment operators for asymmetric top molecules
in case of resonance interactions. The number of interaction parameters
grows very rapidly as the excitation increases. At high energy ranges
which become experimentally accessible new inter-polyad resonances
can occur requiring specific terms to be included in theoretical models.
A. Barbe and co-workers \citep{Barbe_2007,Barbe_2011} have shown
the crucial role of inter-polyad resonance perturbations and of perturbations
due to <<dark>> (experimentally invisible) states, in particular
for ozone specroscopy in the range approaching the dissociation energy.
The probability of near-coincidences of line frequencies rapidely
increases with the number of atoms. For molecules of high-symmetry
the combined effect of degeneracies and resonances makes it nessesary
to model a large number of strongly interacting bands simultanously
\citep{Champion_1992,Nikitin_2002,Albert_2009}. The choice of initial
values for interaction parameters becomes arduous and the minimization
process often reaches a local rather than global minimum. Another
reason is a large proportion of dark states in complex and highly
exited band systems and the corresponding deficit of experimental
information. To some extent, accounting for unobserved energy levels
is equivalent to introducing additional poorly defined parameters
in effective models. On the other hand, full nuclear hamiltonians
expressed in terms of normal-mode irreducible tensor operators can
be obtained from \textit{ab initio} potential energy surfaces (PES)
as reported by Rey et al. \citep{Rey_2010,Rey_2011}. Then, non-empirical
effective hamiltonians can be obtained using high-order contact transformations
\citep{Tyuterev_2004b,Tyuterev_2011,Lamouroux_2008a,Cassam-Chenai_2011}
from a full nuclear hamiltonian. A large proportion of non-diagonal
resonance parameters as well as diagonal parameters for dark states
can be fixed to theoretical values derived from such transformations.
Even though \textit{ab initio} derived values for non diagonal parameters
are approximate, they can describe the resonance couplings in a qualitatively
correct way. 

\textit{Ab initio} vibration-rotation hamiltonians can be converted
into irreducible tensor expansion \citep{Rey_2010,Rey_2011} which
is formally similar to that employed for empirical effective hamiltonians
\citep{Champion_1992,Nikitin_1997a,Zhilinskii_1987} but interestingly
they can be constructed up to higher order than in pure empirical
approaches and are especially suitable for very highly exited band
systems. Of course, the handling of ITO hamiltonian expansions built
from PES relies on advanced computational efficiency. The performance
of programming is important because of the very large number of hamiltonian
terms and basis functions to deal with. For the validation of calculations
it is necessary to check the convergence of contact transformations
\citep{Tyuterev_2004b,Tyuterev_2011,Cassam-Chenai_2011} by achieving
variational calculations with full nuclear hamiltonians. Such calculations
are very cumbersome and require appropriate computational optimization.
The development of efficient computational tools was crucial for the
automatic generation and handling of various types of symmetry allowed
terms simplifying in turn spectral \foreignlanguage{american}{analyses}
and spectroscopic data reductions.

The reader is referred to the original MIRS description \citep{Nikitin_2003a}
and to previous review articles for details on the effective hamiltonian
tensorial approach and results \citep{Champion_1992,Nikitin_1997a,Boudon_2004}.
The present paper is focused on the new computational features, illustrated
by the results obtained so far.

\section{Modeling principles}

\subsection{Irreducible tensor method }

The irreducible tensor formalism and its computer implementation in
MIRS code was already described in details in \citep{Nikitin_1997a,Nikitin_2003a}
and references therein. Let us just remind that within the ITO formalism
each elementary vibration-rotation term 

\begin{equation}
h_{vib-rot}=t_{n_{1}n_{2}...m_{1}m_{2}...}^{\Omega_{r}(K.kC)}\mathbf{\mathit{\mathbf{T}}}_{n_{1}n_{2}...m_{1}m_{2}...}^{\Omega_{r}(K.kC)}=t_{n_{1}n_{2}...m_{1}m_{2}...}^{\Omega_{r}(K.kC)}\mathbf{\mathit{\mathbf{\beta({}^{\varepsilon}V}}}_{n_{1}n_{2}...m_{1}m_{2}...}\mathbf{\mathbf{\otimes R}^{\Omega_{r}(\mathit{K.kC})}})^{A_{1}}
\end{equation}

involved in the hamiltonian expansion is identified by rotational,
vibrational and symmetry indices \citep{CHAMPION_1977} according
to the general nomenclature, where \textbf{V, R }and\textbf{ T} designate
a vibrational, rotational and vibration-rotation tensor operators
and \textit{t }the corresponding parameters. Vibrational operators
$^{\varepsilon}\mathbf{V}_{n_{1}n_{2}...m_{1}m_{2}...}$ are constructed
in tensor form by recursive coupling of creation and annihilation
operators associated to the normal modes of the molecule. The lower
indices \textit{n}$_{i}$ and \textit{m}$_{i}$ being, respectively,
the powers of creation and annihilation vibrational operators. The
upper indices indicate the rotational characteristics of the considered
term: $\Omega{}_{r}$ is the rotational power with respect to the
angular momentum components; \textit{K} is the tensor rank in the
full rotation group; \textit{C} is the rotational symmetry coinciding
with the vibrational symmetry to satisfy the invariance condition
under the molecular point group operations; $\beta$ is a normalization
constant introduced historically \citep{CHAMPION_1977} to ensure
the coincidence between some of the lower order standard and tensorial
spectroscopic constants. Other notations follow references \citep{Champion_1992,Nikitin_1997a,Nikitin_2003a}.

In MIRS, a specific coupling scheme \citep{Nikitin_1997a} associated
to a binary tree is applied directly to an arbitrary number of interacting
vibrational modes and to arbitrarily high polyads. The construction
of vibrational basis functions is achieved consistently from the action
of creation operators on the vacuum function. This method is not only
satisfying conceptually but also quite efficient for the computer
calculation of matrix elements and commutators. The ro-vibrational
terms are easily generated by further tensor couplings. The standard
Amat and Nielsen classification scheme \citep{Amat_1971} is used
in which the order of terms of the type H$_{mn}$ is $m+n-2$. Effective
dipole moment (or other molecular property ) expansions are represented
in a similar manner as that of the hamiltonian but the full symmetry
type has to be appropriately specified according to the molecular
symmetry group \citep{Champion_1992,Nikitin_1997a,Zhilinskii_1987}.
For transformed dipole transition moments, terms of the type $\mu_{mn}$
are of order $m+n-1$.

\subsection{Inter-polyad couplings. Effective and full ro-vibrational hamiltonians}

The effective model definition and associated quantum numbers are
described in detail in Section 2.2 of ref. \citep{Nikitin_2003a}.
Only the basic features are summarized here in order to make the description
of the new implementation understandable. 

The polyad structure constitutes the key for defining effective hamiltonians.
For a given molecule, it is essentially governed by the number of
the vibrational modes and their fundamental frequencies. In the most
straightforward way the polyad scheme is build for vibrational \textit{normal
modes} coupled by vibration-rotation resonance interactions. As a
preliminary step a reference energy level pattern is determined by
resonance conditions defined by the user or by harmonic oscillator
energies computed from the fundamental frequencies. Of course, the
strength of a resonance perturbation depends not only on the proximity
of the zero order energy levels but also on the coupling matrix elements.
The MIRS code allows a certain flexibility for the polyad composition
by optionally including or excluding some vibrational states according
to estimations of coupling matrix elements. It is possible to take
into account such physical considerations by playing with the resonance
conditions. For instance, a gap between reference levels can be artificially
increased in the case of weak vibration-rotation interactions or decreased
in the case of strongly coupled but distant vibrational states. The
true band centers remain unchanged under this operation. According
to this approach the polyad structure of the molecule determines an
effective hamiltonian expansion automatically implemented through
the formal polyad expansion (see \citep{Champion_1992} for details)

\begin{equation}
H^{polyads}=H_{{P_{0}}}+H_{{P_{1}}}+H_{{P_{2}}}+\ldots\;.
\end{equation}

Vibrational matrix elements of the hamiltonian are computed in the
harmonic oscillator zero-order approximation. The computation of rotational
matrix elements is described in \citep{Moret-Bailly_1961,Moret-Bailly_1965,Zhilinskii_1981a}.
The irreducible tensor form \citep{Zhilinskii_1987,Champion_1992,Nikitin_1997a,Boudon_2004}
implemented in the code allows a full account of symmetry properties.
The polyad structure also determines the construction of the effective
transition moments. It determines as well the quantum numbers used
to label levels and transitions. Strict quantum numbers are related
to the usual invariants : the polyad number $P$, the rotational quantum
number $J$ and the ro-vibrational symmetry species $C$.

It is well known that at a low energy range the polyad structure based
on fundamental frequencies usually results in a good approximation
for effective hamiltonians of semi-rigid molecules. For higher quantum
numbers the modeling of experimental data can be considerably improved
by including inter-polyad interactions. In the new version, all features
concerning intra-polyad terms are conserved but inter-polyad terms
can be added. More precisely any inter-polyad operator with rotational
tensor powers equal to zero, one or two may be accounted for. This
new feature may formally be expressed as

\begin{equation}
H=H^{polyads}+H^{inter-polyads}
\end{equation}

\begin{equation}
H^{inter-polyads}=\sum\, H_{{P_{i}/P_{j}}}.
\end{equation}

In this case, the polyad number $P$ is no longer a strict quantum
number. Approximate quantum numbers are generated using the eigenvector
analysis. These approximate quantum numbers should correspond to quantum
numbers of the zero-order harmonic-oscillator + rigid-rotor approximation
under the condition that the corresponding physical quantities are
kept near-invariant under the effect of perturbations. 

In addition to effective ITO spectroscopic models, the new MIRS version
offers a possibility of variational computation of energy levels and
wave functions from a full hamiltonian of nuclear motion in the ITO
representation using symmetrized harmonic oscillator / rigid rotor
basis set for a singlet electronic state. The tools of the MIRS code
allow the generation of all necessary ITO terms up to a requested
order of the expansions (1-4). All necessary matrix elements are calculated
analytically without loss of precision. The values of \textit{t-}parameters
involved in these expansions have to be provided by the user as the
input file for the program. These values can be computed from a given
molecular \textit{ab initio} or empirical PES according to the algorithm
described in our recent works \citep{Rey_2010,Rey_2011} up to arbitrary
orders. At a given order of expansion a full H$_{vib-rot}$ contains
all non-zero symmetry allowed inter-polyad blocks in Eq.(4) systematically
generated from an ab initio PES \citep{Rey_2010,Rey_2011}. Consequently
this procedure usually accounts for many more vibrational interactions
and the full hamiltonian contains more elementary ITO terms than an
effective polyad model. Formally this corresponds to a summation over
all inter-polyad couplings in Eq.(4). But contrary to an effective
spectroscopic model all terms with rotational powers $\Omega{}_{r}$>
2 in the full hamiltonian expansion vanish by definition \citep{Watson_1968}.

\section{Updated database}

The new version of the MIRS program contains a larger number of examples
than the original version \citep{Nikitin_2003a}. The package includes
tutorial examples as well as complete projects related to recent analyses.
New bands were added for CH$_{3}$D, $^{35}$CH$_{3}$Cl, $^{37}$CH$_{3}$Cl
molecules. For CH$_{4}$ and PH$_{3}$ demonstrative examples are
replaced by numerically correct models. A demonstrative model is also
included for CH$_{3}$F based on a full ro-vibrational hamiltonian
(Table \ref{tab:Data-available}). 

\begin{table}
\caption{Data available from the MIRS package\label{tab:Data-available}}
\medskip{}

\centering{}%
\begin{tabular}{ccc}
\hline 
{\footnotesize Molecules} & {\footnotesize Spectral region} & {\footnotesize References}\tabularnewline
\hline 
\hline 
{\footnotesize CH$_{3}$D} & 1500-3700 & \citep{Nikitin_2006}\tabularnewline
\hline 
{\footnotesize CH$_{4}$} & 1700-4600 & \citep{Albert_2009}\tabularnewline
\hline 
{\footnotesize $^{35}$CH$_{3}$Cl} & 0-2600 & \citep{Nikitin_2005a}\tabularnewline
\hline 
{\footnotesize $^{37}$CH$_{3}$Cl} & 0-2600 & \citep{Nikitin_2005a}\tabularnewline
\hline 
{\footnotesize PH$_{3}$} & 700-3500 & \citep{Nikitin_2009}\tabularnewline
\hline 
{\footnotesize CH$_{3}$F} & Global & \tabularnewline
\hline 
\end{tabular}
\end{table}

A quite simple example is the sixth-order model for the ground-state,
dyad, pentad and octad of $^{12}$CH$_{4}$. In this case, the vibrational
coupling scheme coincides exactly with the one used in the STDS program
\citep{Wenger_1998}. This means that the effective parameter sets
are identical in MIRS and STDS providing a good validation of both
programs. For the following polyads (from the tetradecad upwards),
even though the models are equivalent and include the same number
of parameters at a given order of approximation, the two parameter
sets are not in one to one correspondence. A similar system is illustrated
with the sixth order model for the ground-state, dyad, pentad and
octad of PH$_{3}$ \citep{Nikitin_2009}.

The model for the lower two polyads (triad and nonad) of CH$_{3}$D
represents a complex band system of a symmetric top molecule \citep{Nikitin_2006}.
The parameters of the MIRS model have been fitted to some 12589 line
positions and to 2400 line intensities with an accuracy close to the
experimental precision (about 0.001 cm$^{-1}$ for positions and 4\%
for intensities). Altogether a common set of 441 effective hamiltonian
parameters has been used to describe 13 vibrational states (ground
state, triad, nonad) corresponding to 39 bands (6 fundamental, 3 overtone,
3 combination, and 27 hot bands). The intensities of the 9 bands involved
in the nonad-ground state system were modeled at the second order
of approximation using 83 effective dipole transition moment parameters.

Finally, results of the global analysis of the lower polyads of $^{12}$CH$_{3}$$^{35}$Cl
and $^{12}$CH$_{3}$$^{37}$Cl are included. A preliminary analysis
of the infrared spectrum in the region from 0 to 2600 cm$^{-1}$ was
performed using 288(303) effective parameters for isotopomers $^{12}$CH$_{3}$$^{35}$Cl($^{12}$CH$_{3}$$^{37}$Cl)
to model five lower polyads (17 vibrational states\textbf{) }\citep{Nikitin_2005a}\textbf{.}
The precision on positions is of the order of 0.0007 cm$^{-1}$.

\section{Applications}

\subsection{Basic applications}

The basic applications already implemented in the original version
\citep{Nikitin_2003a} are maintained in the new one. Complete files
of energy levels can be calculated using the command Calculus > build.
Each record contains the energy value, the set of rigorous quantum
numbers (total angular momentum $J$, symmetry point group irreducible
representations and ranking number) as well as approximate quantum
numbers derived from the coefficients of the eigenvector expansions
in the initial basis set. Details on the experimental measurements
fitted are also provided. The number and the root-mean-squares (rms)
deviations of assigned transitions are given together with the corresponding
averaged observed minus calculated residual for every example. The
partition function can be derived from the calculated levels involved
in the model or set to an appropriate external estimate \citep{Wenger_2008a}.
Several output formats including those of HITRAN \citep{Rothman_2009}
and GEISA \citep{Jacquinet-Husson_2008} can be selected. The temperature
and intensity threshold can be adjusted to meet specific needs for
applications.

\subsection{Advanced applications}

\subsubsection{Setting up new models}

\begin{figure}
\caption{MIRS screen copy showing the model definition for CH$_{3}$Cl \label{fig:mirs5}}
\medskip{}

\begin{centering}
\includegraphics[width=15cm]{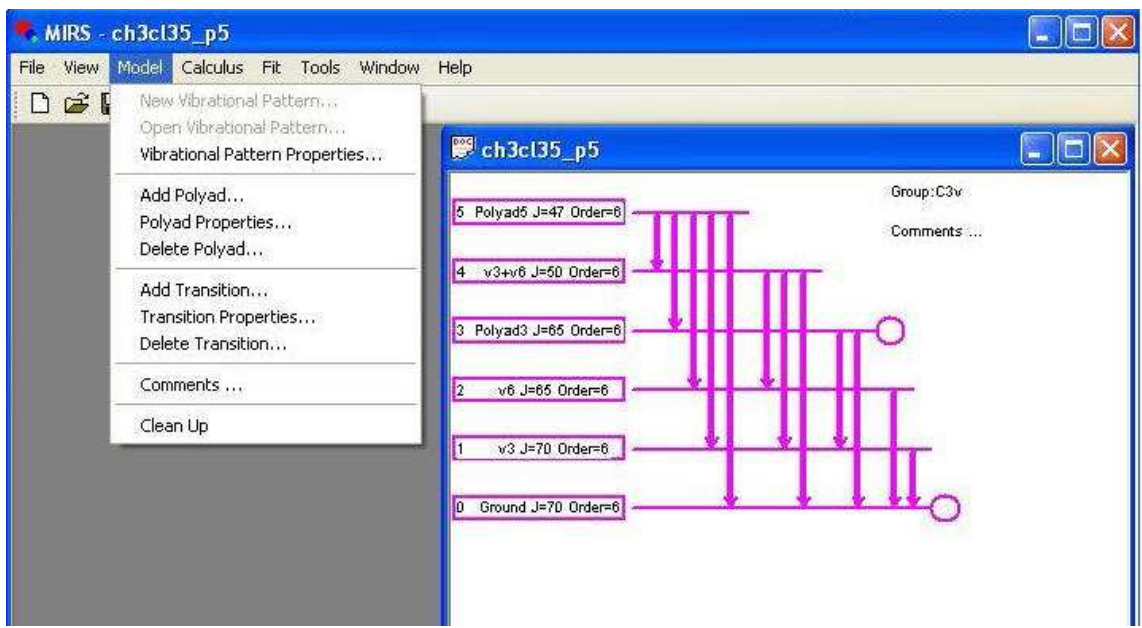}
\par\end{centering}

\medskip{}

{\footnotesize This model includes the ground state and the lower
five polyads of CH$_{3}$Cl. The arrows represent the band systems
involved : 5 originating from the ground base, 3 from the $\nu_{3}$
band, 2 from the $\nu_{6}$ band, 1 from polyad 3 and one from the
$\nu_{3}+\nu_{6}$ band. The circles represent pure rotational transitions
within the ground state and polyad 3 upper states. All the corresponding
observed data (energy levels, line positions and line intensities)
can be fitted coherently.}\medskip{}
\end{figure}

In order to set up a new model it is necessary to define the polyad
structure and to introduce the modeling parameters. The modeling parameters
are: (i) the symmetry point group of the molecule, (ii) the harmonic
frequencies of the considered normal modes, (iii) the corresponding
symmetry species (irreducible representations), (iv) a frequency gap
used for gathering the subsequent vibrational sublevels into relevant
vibrational polyads. Note that in the modeling procedure the frequencies
can be entered in arbitrary units and do not need to match exactly
the physical frequencies in order to provide flexibility. For example,
the modeling of the CH$_{3}$D molecule was performed by setting the
six fundamental frequencies to 2, 2, 1, 2, 1, 1 respectively and the
frequency gap to 0.1. This yielded the triad and nonad structure described
in \citep{Nikitin_2002,Nikitin_2006}. For CH$_{3}$Cl the situation
was more complex and the fundamental frequencies were set closer to
the actual values: 2968, 1354, 733, 3039, 1452 and 1018, respectively
with a frequency gap of 200 (Fig. \ref{fig:mirs5}) \citep{Nikitin_2003a,Nikitin_2005a,Nikitin_2005}.
This method of the model set up is quite general. For example choosing
a large energy gap results automatically in including a maximum of
symmetry allowed vibration-rotation interaction terms. Then by removing
some of them (via text-editor) one can build a model suitable for
describing observed resonance perturbations in molecular spectra.
By this way one can easily build an \textit{<<overlapping-polyad>>
model} in order to account for inter-polyad resonances. In all cases,
the resulting polyad structure can be displayed for checking and modifying
purposes. 

By default, MIRS is set to build effective hamiltonian matrices and
transition moments according to the vibrational extrapolation scheme
in the normal mode harmonic oscillator basis. This means that for
a given molecule a common set of effective hamiltonian parameters
is set up to fit and predict the subsequent vibrational polyads. The
same scheme applies for transition moment parameters. For instance
a band system from the ground state to a given polyad and all the
corresponding hot band systems are described by a common set of effective
parameters. Such features are generated automatically by the program
and controlled by a few basic modeling parameters entered through
the window menu reproduced in Fig. \ref{fig:mirs5}. 

\begin{figure}
\caption{MIRS option window \label{fig:options}}
\medskip{}

\begin{centering}
\includegraphics[width=15cm]{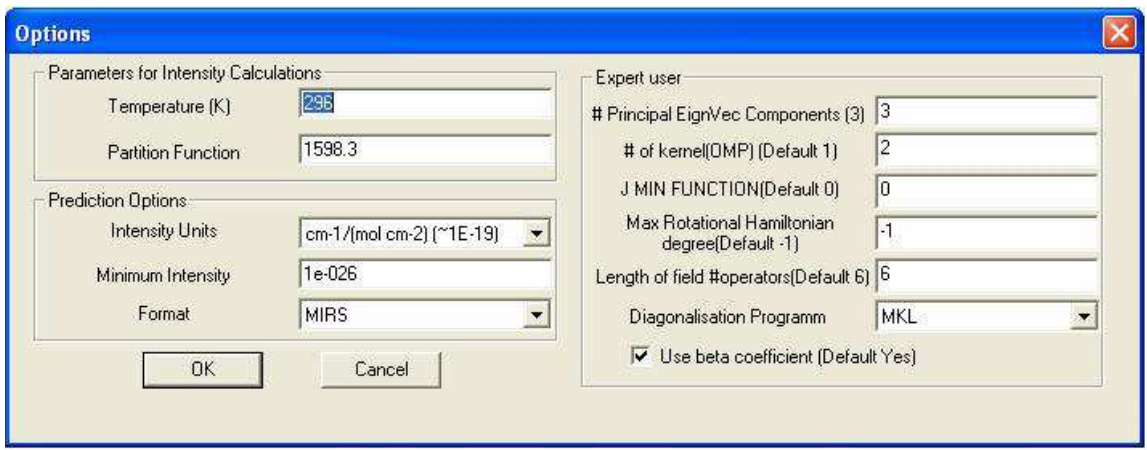}
\par\end{centering}

\medskip{}

\centering{}{\footnotesize Refer to the Web site}{\scriptsize{} }{\footnotesize for
explanations about non trivial options }\medskip{}
\end{figure}

The nuclear spin statistical weights involved in intensity calculations
are also manually introduced at this stage. Once the vibrational model
is set up, the polyads and the transitions to consider are entered
through another window menu exemplified in Fig. \ref{fig:options}.
For each polyad the order of the hamiltonian expansion and the maximum
value of $J$ has to be entered. Similarly, for each type of transitions
the order of the effective dipole moment expansion has to be specified.

At the end of the procedure the build command generates all internal
files needed for subsequent calculations. The same command is used
to update all files whenever needed. The other commands of the Calculus
menu displayed in Fig. \ref{fig:options} are generally self-explanatory.
More details are available from the Help command. Thanks to its intrinsic
flexibility MIRS can also be used to create quantum model hamiltonians
not necessarily connected to real molecules. For instance, various
systems involving coupled harmonic or anharmonic oscillators can be
simulated numerically for further theoretical investigations.

\subsubsection{Experimental data fitting (positions and intensities)}

MIRS offers extensive set of tools for a non-linear least-squares
fit of spectroscopic data (positions and intensities). According to
the vibrational extrapolation scheme \citep{Champion_1992} well suited
for global analyses, the program is set to fit simultaneously all
transitions and/or vibration-rotation levels involved in a given project.
The fit of intensities is also achieved using a non-linear least-squares
procedure with an adequate weighting already described in \citep{Wenger_1998,Wenger_2000}.
Note that in our approach a rigorous consistency is de facto applied
between intensity and position fits. The reader is referred to the
original paper \citep{Nikitin_2003a} for details on the sophisticated
algorithms developed to overcome the increasing complexity of the
problems to solve. 

In the new version, two subprograms have been added to handle the
adjustable parameters and their possible constraints. The Tools menu
includes the Free\_Parameters item to release all parameters that
were previously fixed automatically by the program. The Copy\_Parameters
item allows the copy of parameters from one file to another. This
is especially useful when one needs to change the model, for example
to increase the order of the hamiltonian expansion.

Another improvement has been added to manage situations encountered
when approximate quantum numbers are not physically meaningful and
well conserved. The assignment of observed to calculated data is based
on the rigorous quantum numbers. For an isolated polyad model these
are: $P$ (polyad number), $J$ (total angular momentum), $C$ (symmetry
species) and the ranking index $n$ within each $P,J,C$ block. In
some cases of nearby eigenvalues the ranking indices may undergo permutations
from one iteration to another resulting to undesirable assignment
jumps. MIRS allows temporarily removing from the fit the assigned
levels which present such behavior by detecting anomalous variations
of the scalar product of the corresponding eigenvectors during the
iteration process. For this one needs to calculate at least parts
of the eigenvector matrix $\alpha$. If $f_{i}^{(0)}$ designates
the $i^{th}$ initial eigenfunction within a given $P,J,C$ block
and $f_{j}^{(k)}$ the $j^{th}$ current eigenfunction within the
same $P,J,C$ block, the above mentioned scalar product can be expressed
as 
\begin{equation}
<f_{j}^{(0)}|f_{i}^{(k)}>=\underset{\sigma\delta}{\sum}\alpha_{j\sigma}^{(0)}\alpha_{i\delta}^{(k)}<\sigma|\delta>=A_{j,i},\label{eq:scalar_product}
\end{equation}

where $\sigma$ and $\delta$ denote basis functions. For the great
majority of the eigenvectors the scalar product (\ref{eq:scalar_product})
is close to unity and equivalently the $A$ matrix is almost diagonal.
In a minority of cases this is no longer true. For example an $A$
matrix of the form

\begin{equation}
A=\left[\begin{array}{cccccc}
\ldots & \ldots & \ldots & \ldots & \ldots & \ldots\\
 & 0.98 & 0.01 & 0.00 & 0.00 & \ldots\\
 &  & 0.20 & 0.97 & 0.14 & \ldots\\
 &  &  & 0.20 & 0.14 & \ldots\\
 &  &  &  & 0.65 & \ldots\\
 &  &  &  &  & \ldots
\end{array}\right]\label{eq:A_matrix}
\end{equation}

illustrates a situation where the two vectors corresponding to the
middle of the matrix are rearranged while the first vector, associated
to the diagonal element equal to 0.98, keeps its initial $n$ index.
Sometimes the rearrangement is not obvious, for instance when the
diagonal element is not close to unity and simultaneously all non-diagonal
values are not very large. This applies for the fourth vector of the
matrix \ref{eq:A_matrix} with diagonal matrix element equal to 0.65.
In this case the corresponding level must be excluded from the fit
at least temporarily. The file with the initial eigenvectors is saved
before the fit. For every iteration, some of the assigned levels are
excluded from the fit and part of the levels previously excluded are
re-assigned. If a large proportion of levels are excluded from the
fit, then the iteration is unsuccessful and it becomes necessary to
decrease the allowed variations for the parameters.

The fit of intensities is also subject to specific problems. Specific
features to handle relative signs between transition parameters and
interaction hamiltonian parameters are already described in section
4.3 of Ref. \citep{Nikitin_2003a}. 

Despite all the above advanced options, it should be emphasized that
in any case a purely automatic selection of adjustable terms for positions
or intensities in an arbitrarily built model is not recommended. All
procedures have to rely on physical considerations and depend on experimental
data set and observed perturbations involved.

\subsubsection{Group theory re-coupling coefficients and commutator-algebra calculations}

The calculations involved in effective hamiltonians for highly exited
states can be also very cumbersome and requires program optimization.
It was found that one of the bottlenecks is the slow calculation of
$9C$ \citep{Champion_1992} (other notations: X \citep{Griffith_1962},
$9\Gamma$ symbols\citep{Zhilinskii_1987}). The $12C$ symbols are
required only for intensity calculations while $9C$ are used for
line position calculations. In addition, $9C$ symbols are used in
variational programs designed for the calculation of multidimensional
irreducible matrix elements \citep{Nikitin_2009a,Nikitin_2008,Nikitin_2011a}.
In the present version of MIRS, the calculation of $6C$, $9C$ and
$12C$ symbols for C$_{3v}$, T$_{d}$, and O$_{h}$ point groups
is based on the improved algorithms reported in \citep{Nikitin_2011b}
which is much more efficient than in the program previously used.
This new algorithm \citep{Nikitin_2011b} uses the symmetry relations
between even and odd representations. It allows one to speed up the
$C$-symbol calculation and increase the efficiency of spectroscopic
programs based on the irreducible tensorial formalism.

MIRS can also be used by expert users to generate various types of
coupling coefficients as well as commutators of ITOs in algebraic
or numerical formats. Such tools can be useful for the theoretical
investigation of the fundamental properties of effective models and
for establishing their relations with the full nuclear hamiltonian.
It is well-known that, in general, an effective Hamiltonian is not
uniquely determined from its eivenvalues \citep{WATSON_1967,KLEIN_1974,Tyuterev_1975,TYUTEREV_1980,Makushkin_1984,JORGENSEN_1975}.
The ambiguity of the determination of $H^{{eff}}$ from experimental
energies is particularly important in the case of resonance interactions
due to quasi-degeneracies of vibrational states \citep{TYUTEREV_1980,PEREVALOV_1982b,PEREVALOV_1984a,TYUTEREV_1986,LOBODENKO_1987a}.
In order to improve the convergence of least-square fits to experimental
spectra and to obtain a unique set of fitted parameters, the spectroscopic
effective hamiltonian models have to be constrained using the so called
<<reduction theory>>\citep{WATSON_1967,Aliev_1985,PEREVALOV_1982b,PEREVALOV_1984,LOBODENKO_1987a,TYUTEREV_1990,Sarka_2000}.
The latter relays on extended commutator / anti-commutator algebra
calculations. This Lie algebra of ITO plays a key role for a full
account of symmetry properties in the contact transformation method
\citep{Makushkin_1984,Zhilinskii_1987,Tyuterev_2002}.

\subsubsection{Generation of symmetry allowed terms in the irreducible tensor form}

We have added into the new MIRS a possibility for the user to generate
a list of symmetry allowed terms involved in the ITO normal mode expansion
of either an effective polyad hamiltonian (including inter-polyad
interactions), or of an effective dipole moment or of the potential
energy function up to an arbitrary order. For example, the list of
ITO terms involved in a second order PES expansion for an XY$_{4}$
molecule can be obtain from the menu <<tools > build parameters>>
by choosing <<type of file = PES>>, then by choosing a vibrational
pattern, and finally by defining the desired model. A similar MIRS
tool is also operational for listing all allowed ITO terms involved
in generators of unitary transformations which preserve group symmetry,
time reversal and hermicity properties of transformed physical properties
of a molecule. The present version of the package includes a sample
example of symmetry allowed terms in S-generators involved in the
problem of the effective hamiltonian reduction for the triad of CH$_{3}$D.

\section{Modified specifications}

\subsection{Vibrational and Ro-vibrational calculation}

All MIRS parameter files, dipole moment files and basis function files
are text files. In order to calculate highly exited states using full
nuclear hamiltonians a new coding of the symmetric powers ($l,m,n$)
\citep{Nikitin_1997a} was necessary. The original version of the
program used only digits $0,..,9$ for triplet numbers ($l,m,n$).
So only symmetric powers up to 9 were available although new applications
required numbers with two digits. To ensure the compatibility with
the original version and to preserve a good legibility for low symmetric
powers, we decided to keep one single symbol according to Table \ref{tab:symbolic-coding}.
Note that now the number of polyads is limited to 99 which is far
beyond the limit of realistic calculations.

\begin{table}
\caption{Symbolic coding of vibrational powers\label{tab:symbolic-coding}}
\medskip{}

\centering{}%
\begin{tabular}{cc}
\hline 
{\footnotesize Numbers} & {\footnotesize Symbols}\tabularnewline
\hline 
\hline 
{\footnotesize $0,1,2,\ldots,9$} & {\footnotesize $0,1,2,\ldots,9$}\tabularnewline
\hline 
{\footnotesize $10,11,12,\ldots,35$} & {\footnotesize $A,B,C,\ldots,Z$}\tabularnewline
\hline 
{\footnotesize $36,37,38,\ldots,61$} & {\footnotesize $a,b,c,\ldots,z$}\tabularnewline
\hline 
{\footnotesize $62,63$} & {\footnotesize $\{,|$}\tabularnewline
\hline 
{\footnotesize $>63$} & ?\tabularnewline
\hline 
\end{tabular}
\end{table}

\subsection{Options}

For efficient calculations on multi-kernel (multiprocessors) computers,
the new MIRS program uses OpenMP version 2.0 (\href{http://openmp.org}{http://openmp.org}).
Up to 8 kernels can be used simultaneously for matrix element calculations.
As mentioned previously the full \foreignlanguage{american}{ro-vibrational}
hamiltonian contains terms up to the second rotational power only,
while effective hamiltonian may contain all possible rotational powers.
Different programs may be used for eigenvalue calculation (see Diagonalization
option). It is sometimes convenient to store on disk the full matrix
for each $P,J,C$ block because the diagonalization of a large number
of matrices may be very burdensome.

\subsection{Limitations}

The previous version of the MIRS code was a 32-bit one with considerable
restrictions on the number of tensors involved regarding the power
of hamiltonian expansions and the size of polyads. The present version
extends the number of polyads $P$ from 9 to 99 while the number of
terms and the polyad size is limited by the available computer memory.
In order to speed-up calculations, sufficient computer resources are
required to store all vibrational matrix elements in random access
memory. The number of molecular vibration modes is limited to 20 and
the maximum total angular momentum quantum number $J$ for the T$_{d}$
point group is limited to 199 \citep{Rey_2003}.

\section{Getting and installing}

The MIRS package can be downloaded freely from our Web sites \href{http://www.iao.ru/mirs/mirs.htm}{http://www.iao.ru/mirs/mirs.htm}
\footnote{also available from \href{http://xeon.univ-reims.fr/Mirs/}{http://xeon.univ-reims.fr/Mirs/}
or \href{http://icb.u-bourgogne.fr/OMR/SMA/SHTDS/MIRS.html}{http://icb.u-bourgogne.fr/OMR/SMA/SHTDS/MIRS.html}%
} and as supplementary data in the online version of the article. It
consists of a self-extractive archive containing various tutorial
examples. MIRS, written in C++, includes all necessary executable
files for Windows XP and higher. Two versions (32 bit and 64 bit)
of executable files are available. MIRS 64 bit works only on 64 bit
windows operating systems. The major part of the examples based on
effective hamiltonians (except CH4\_polyad7 and CH4\_polyad9) can
be run using the 32 bit version. Fit operations can be run on the
32 bit version only. As a rule, full hamiltonian examples can be calculated
using the 64 bit version only. Memory, CPU and disk space requirements
depend on the complexity of the project. The package itself requires
a minimum of 15 MB of free disk space. The tables of coupling coefficients
occupy up to 150 MB when the highest $J$ value considered so far
(80) is involved. All examples and projects included in the package
can be executed using a computer with at least 2 GB RAM and 500 MB
of free disk space. A typical execution time using a processor at
2 GHz is about 10 min to build the CH$_{3}$D project (all matrix
elements and prediction files) and less than 15 min to build each
of the CH$_{3}$Cl projects. System requirements for complex examples
are described in readme.txt files. For example CH4\_polyad9 can be
calculated only on computer with at least 24GB of memory and 160GB
of free disk space. Standard features and options are directly accessible
from the tutorial projects in which all necessary modeling files are
included. The Help menu can be accessed at any stage of the project
building and subsequent calculations. Refer to our Web sites for complete
installation and running instructions.

\section{Conclusion}

The rapid increase of computer power and the progresses in multiprocessor
parallellization have opened the way to considerable improvements
of theoretical predictions and data reduction for complex high-energy
band systems of polyatomic molecules. This may help in understanding
new spectra involving high vibration-rotation excitations which have
become experimentally accessible recently or can be available in future.
Efficient modeling of spectra requires the development of advanced
software specifically optimized for these purposes. Various methods
and program implementations have been developed for computation of
molecular spectra either from effective spectroscopic models \citep{Tashkun_2010,Sarka_2000,Wenger_1998,Champion_1992,FLAUD_1975,Tashkun_1993,Herman_2007,Amyay_2011}
(and references therein) or from potential and dipole moment surfaces
using variational approaches \citep{JENSEN_1988d,JENSEN_1988e,Schwenke_1996,Bowman_2010,Yurchenko_2011,CARTER_1986,Nikitin_2011c,Zuniga_2007,Schwenke_2002,Matyus_2009,Tyuterev_2004},
discrete variable representation (DVR) \citep{Light_2000,Wang_2011b,Wang_2011c,Tennyson_2000,Tennyson_1992,Wang_2002}
or filter-digitalization \citep{Mandelshtam_1997} techniques. A comparative
analysis of advantages and limitations of these approaches is beyond
the scope of the present paper ( see ref. \citep{Tyuterev_2003a}
for some related discussion). Complementary results for non-empirical
effective hamiltonians and molecular spectra can be obtained via perturbation
theory and Contact Transformation methods \citep{VanVleck_1929,Amat_1971,Aliev_1985,TYUTEREV_1980,Makushkin_1984,SIBERT_1988a,Camy-Peyret_1985,Wang_1999,Sarka_2000,Tyuterev_2004b,Lamouroux_2008a,Cassam-Chenai_2011}.
An efficient computational software for the experimental data reduction
in molecular spectroscopy and for the generation of line list parameters
allows completing and extending spectroscopic databases which are
planned to be linked in frame of the <<Virtual Atomic and Molecular
Data Center>> (VAMDC) European project \citep{Dubernet_2010}. 

The reported extension of the MIRS computer package is designed to
fully account for symmetry properties by using irreducible tensor
representations for all physical properties and operators involved
in direct calculations as well as in experimental molecular spectra
fitting. Its particular feature is a unified treatment of effective
polyad models (with the possibility of including explicitly inter-polyad
interactions) and of full ab initio hamiltonian normal mode expansions
in the frame of the ITO formalism. The new MIRS includes more efficient
and faster algorithms for the calculation of coefficients related
to molecular symmetry properties ($6C$, $9C$ and $12C$ symbols
for C$_{3v}$, T$_{d}$, and O$_{h}$ point groups) and for a better
convergence of least-square-fit iterations. Several user-friendly
tools are included dealing with the ITO algebra and with the generation
of symmetry allowed terms in spectroscopic effective hamiltonians,
transition moment operators, potential energy function, full ab initio
hamiltonian expansions, and unitary transformations involved in the
data reduction theory. This approach combined with an appropriate
implementation of the high-order Contact Transformation method \citep{Tyuterev_2004b,Tyuterev_2011}
can be used, in principle, to perform a systematic computation of
ab initio values of resonance coupling parameters in spectroscopic
hamiltonians which were poorly determined in purely empirical models.
The derivation of spectroscopic models for the CH$_{3}$Cl, CH$_{3}$Br,
CH$_{3}$I, GeH$_{4}$, AsH$_{3}$ molecules constructed in this way
is currently in progress. We plan to provide in future a UNIX/LINUX
version of MIRS.

\section*{Acknowledgments}

Part of the work was performed in the framework of LEFE-CHAT National
Program CNRS (France). Support from Agence Nationale de la Recherche
(France) through the project CH4@Titan (ref: BLAN08- 2 321467) , from
GDRI SAMIA involving Tomsk (Russia), Hefei (China), CNRS (France)
and from the VAMDC EC project is gratefully acknowledged.

\section*{Installation package}

The installation package is freely available from the attached file.

\newpage{}

\bibliographystyle{elsarticle-num}
\phantomsection\addcontentsline{toc}{section}{\refname}\bibliography{biblio_mirs_2011}

\begin{thebibliography}{10}
\expandafter\ifx\csname url\endcsname\relax
  \def\url#1{\texttt{#1}}\fi
\expandafter\ifx\csname urlprefix\endcsname\relax\def\urlprefix{URL }\fi
\expandafter\ifx\csname href\endcsname\relax
  \def\href#1#2{#2} \def\path#1{#1}\fi

\bibitem{Champion_1992}
J.~P. Champion, M.~Loete, G.~Pierre, Spherical top spectra, in: K.~Rao,
  A.~Weber (Eds.), Spectroscopy of the Earth's atmosphere and interstellar
  medium, Academic Press Inc., Columbus, 1992, pp. 339--422.

\bibitem{Hilico_2001}
J.~C. Hilico, O.~Robert, M.~Loete, S.~Toumi, A.~S. Pine, L.~R. Brown, Analysis
  of the interacting octad system of $^{12}${CH}$_{4}$, J Mol Spectrosc 208
  (2001) 1--13.
\newblock \href {http://dx.doi.org/10.1006/jmsp.2001.8364}
  {\path{doi:10.1006/jmsp.2001.8364}}.

\bibitem{Rotger_2000a}
M.~Rotger, V.~Boudon, M.~Loete, Spectroscopy of {XY}$_5${Z} {C}$_{4v}$
  molecules: Development of the hamiltonian and the transition moment operators
  using a tensorial formalism, J Mol Spectrosc 200 (2000) 131--137.
\newblock \href {http://dx.doi.org/10.1006/jmsp.1999.8035}
  {\path{doi:10.1006/jmsp.1999.8035}}.

\bibitem{Nikitin_2003a}
A.~Nikitin, J.~P. Champion, V.~G. Tyuterev, The {MIRS} computer package for
  modeling the rovibrational spectra of polyatomic molecules, JQSRT 82 (2003)
  239--249.
\newblock \href {http://dx.doi.org/10.1016/S0022-4073(03)00156-0}
  {\path{doi:10.1016/S0022-4073(03)00156-0}}.

\bibitem{Nikitin_2004}
A.~Nikitin, J.~P. Champion, H.~Burger, Global analysis of chloromethane:
  determinability of ground state constants, Vol. 5311, Proc. SPIE, 2004, pp.
  97--101.
\newblock \href {http://dx.doi.org/10.1117/12.545197}
  {\path{doi:10.1117/12.545197}}.

\bibitem{Boudon_2004}
V.~Boudon, J.~P. Champion, T.~Gabard, M.~Loete, F.~Michelot, G.~Pierre,
  M.~Rotger, C.~Wenger, M.~Rey, Symmetry-adapted tensorial formalism to model
  rovibrational and rovibronic spectra of molecules pertaining to various point
  groups, J Mol Spectrosc 228 (2004) 620--634.
\newblock \href {http://dx.doi.org/10.1016/j.jms.2004.02.022}
  {\path{doi:10.1016/j.jms.2004.02.022}}.

\bibitem{Nikitin_2005a}
A.~Nikitin, J.~P. Champion, H.~Burger, Global analysis of
  $^{12}${CH}$_{3}$$^{35}${C}l and $^{12}${CH}$_{3}$$^{37}${C}l: simultaneous
  fit of the lower five polyads (0-2600 cm$^{-1}$), J Mol Spectrosc 230 (2005)
  174--184.
\newblock \href {http://dx.doi.org/10.1016/j.jms.2004.11.012}
  {\path{doi:10.1016/j.jms.2004.11.012}}.

\bibitem{Nikitin_2006}
A.~V. Nikitin, J.~P. Champion, L.~R. Brown, Preliminary analysis of {CH}$_3${D}
  from 3250 to 3700 cm$^{-1}$, J Mol Spectrosc 240 (2006) 14--25.
\newblock \href {http://dx.doi.org/10.1016/j.jms.2006.08.002}
  {\path{doi:10.1016/j.jms.2006.08.002}}.

\bibitem{Rotger_2002}
M.~Rotger, V.~Boudon, M.~Loete, Spectroscopy of {XY}$_2${Z}$_2$ {C}$_{2v}$
  molecules: A tensorial formalism adapted to the {O}(3) $\supset$ {T}$_{d}$
  $\supset$ {C}$_{2v}$ chain. application to the ground state of
  {SO}$_2${F}$_2$, J Mol Spectrosc 216 (2002) 297--307.
\newblock \href {http://dx.doi.org/10.1006/jmsp.2002.8635}
  {\path{doi:10.1006/jmsp.2002.8635}}.

\bibitem{Nikitin_2002}
A.~Nikitin, L.~R. Brown, L.~Fejard, J.~P. Champion, V.~G. Tyuterev, Analysis of
  the {CH}$_{3}${D} nonad from 2000 to 3300 cm$^{-1}$, J Mol Spectrosc 216
  (2002) 225--251.
\newblock \href {http://dx.doi.org/10.1006/jmsp.2002.8566}
  {\path{doi:10.1006/jmsp.2002.8566}}.

\bibitem{Albert_2009}
S.~Albert, S.~Bauerecker, V.~Boudon, L.~R. Brown, J.~P. Champion, M.~Loete,
  A.~Nikitin, M.~Quack, Global analysis of the high resolution infrared
  spectrum of methane $^{12}${CH}$_{4}$ in the region from 0 to 4800 cm$^{-1}$,
  Chem Phys 356 (2009) 131--146.
\newblock \href {http://dx.doi.org/10.1016/j.chemphys.2008.10.019}
  {\path{doi:10.1016/j.chemphys.2008.10.019}}.

\bibitem{FLAUD_1975}
J.~M. Flaud, C.~Camy-Peyret, Vibration-rotation intensities in {H}$_2${O}-type
  molecules application to $2\nu_2$-band, $\nu_1$-band, and $\nu_3$-band of
  {H}$_2^{16}${O}, J Mol Spectrosc 55 (1975) 278--310.
\newblock \href {http://dx.doi.org/10.1016/0022-2852(75)90270-2}
  {\path{doi:10.1016/0022-2852(75)90270-2}}.

\bibitem{Camy-Peyret_1985}
C.~Camy-Peyret, J.~M. Flaud, Vibration-rotation dipole moment operator for
  asymmetric top molecules, in: K.~N. Rao (Ed.), Molecular Spectroscopy: Modern
  Reseach, Vol.~3, Acadmic Press, Orlando, FL, 1985, pp. 69--109.

\bibitem{Barbe_2007}
A.~Barbe, M.~R. De~Backer-Barilly, V.~G. Tyuterev, A.~Campargue, D.~Romanini,
  S.~Kassi, {CW}-cavity ring down spectroscop, of the ozone molecule in the
  5980-6220 cm$^{-1}$ region, J Mol Spectrosc 242 (2007) 156--175.
\newblock \href {http://dx.doi.org/10.1016/j.jms.2007.02.022}
  {\path{doi:10.1016/j.jms.2007.02.022}}.

\bibitem{Barbe_2011}
A.~Barbe, M.~R. De~Backer-Barilly, V.~G. Tyuterev, S.~Kassi, A.~Campargue,
  Detection and analysis of new bands of $^{16}${O}$_3$ by {CRDS} between 6500
  and 7300 cm$^{-1}$, J Mol Spectrosc 269 (2011) 175--186.
\newblock \href {http://dx.doi.org/10.1016/j.jms.2011.06.005}
  {\path{doi:10.1016/j.jms.2011.06.005}}.

\bibitem{Rey_2010}
M.~Rey, A.~V. Nikitin, V.~G. Tyuterev, Ab initio ro-vibrational hamiltonian in
  irreducible tensor formalism: a method for computing energy levels from
  potential energy surfaces for symmetric-top molecules, Mol Phys 108 (2010)
  2121--2135.
\newblock \href {http://dx.doi.org/10.1080/00268976.2010.506892}
  {\path{doi:10.1080/00268976.2010.506892}}.

\bibitem{Rey_2011}
M.~Rey, A.~V. Nikitin, V.~G. Tyuterev, The complete nuclear hamiltonian in the
  irreducible tensor operator formalism for methane, to be published.

\bibitem{Tyuterev_2004b}
V.~G. Tyuterev, S.~A. Tashkun, H.~Seghir, High-order contact transformations:
  General algorithm, computer implementation and triatomic tests, Vol. 5311,
  Proc. SPIE, 2004, pp. 164--175.
\newblock \href {http://dx.doi.org/10.1117/12.545641}
  {\path{doi:10.1117/12.545641}}.

\bibitem{Tyuterev_2011}
V.~G. Tyuterev, S.~A. Tashkun, M.~Rey, A.~V. Nikitin, R.~Rochanov, High-order
  contact transformations for methane molecule, in preparation.

\bibitem{Lamouroux_2008a}
J.~Lamouroux, S.~A. Tashkun, V.~G. Tyuterev, Accurate calculation of transition
  moment parameters for rovibrational bands from ab initio dipole and potential
  surfaces: Application to fundamental bands of the water molecule, Chem Phys
  Lett 452 (2008) 225--231.
\newblock \href {http://dx.doi.org/10.1016/j.cplett.2007.12.061}
  {\path{doi:10.1016/j.cplett.2007.12.061}}.

\bibitem{Cassam-Chenai_2011}
P.~Cassam-Chenai, Y.~Bouret, M.~Rey, S.~Tashkun, A.~V. Nikitin, V.~G. Tyuterev,
  Ab initio effective rotational hamiltonians: A comparative study, Int J
  Quantum Chem Article published online.
\newblock \href {http://dx.doi.org/10.1002/qua.23183}
  {\path{doi:10.1002/qua.23183}}.

\bibitem{Nikitin_1997a}
A.~Nikitin, J.~P. Champion, V.~G. Tyuterev, Improved algorithms for the
  modeling of vibrational polyads of polyatomic molecules: Application to
  {T}$_d$, {O}$_h$, and {C}$_{3v}$ molecules, J Mol Spectrosc 182 (1997)
  72--84.
\newblock \href {http://dx.doi.org/10.1006/jmsp.1996.7185}
  {\path{doi:10.1006/jmsp.1996.7185}}.

\bibitem{Zhilinskii_1987}
B.~Zhilinskii, V.~Perevalov, V.~G. Tyuterev, Method of irreducible tensorial
  operators in the theory of molecular spectra. (in Russian), Nauka,
  Novosibirsk, 1987.

\bibitem{CHAMPION_1977}
J.~P. Champion, Complete development of vibration-rotation hamiltonian adapted
  to study of interactions in spherical top molecules - application to $\nu_2$
  and $\nu_4$ bands of $^{12}${CH}$_{4}$, Can J Phys 55 (1977) 1802--1828.
\newblock \href {http://dx.doi.org/10.1139/p77-221}
  {\path{doi:10.1139/p77-221}}.

\bibitem{Amat_1971}
G.~Amat, H.~H. Nielsen, G.~Tarrago, Rotation - vibration of polyatomic
  molecules, Dekker, New York, 1971.

\bibitem{Moret-Bailly_1961}
J.~Moret-Bailly, Cah Phys 15 (1961) 237--314.

\bibitem{Moret-Bailly_1965}
J.~Moret-Bailly, Calculation of the frequencies of the lines in a threefold
  degenerate fundamental band of a spherical top molecule, J Mol Spectrosc 15
  (1965) 344--354.

\bibitem{Zhilinskii_1981a}
B.~Zhilinskii, Reduction of rotational operators to standard form, Opt
  Spectrosc (USSR) 51~(3) (1981) 262--263.

\bibitem{Watson_1968}
J.~K.~G. Watson, Simplification of molecular vibration-rotation hamiltonian,
  Mol Phys 15 (1968) 479--490.
\newblock \href {http://dx.doi.org/10.1080/00268976800101381}
  {\path{doi:10.1080/00268976800101381}}.

\bibitem{Nikitin_2009}
A.~V. Nikitin, Modeling of vibrational energy levels of methane from the ab
  initio constructed potential energy surface, Optics and Spectroscopy 106
  (2009) 176--182.
\newblock \href {http://dx.doi.org/10.1134/S0030400X09020052}
  {\path{doi:10.1134/S0030400X09020052}}.

\bibitem{Wenger_1998}
C.~Wenger, J.~P. Champion, Spherical top data system ({STDS}) software for the
  simulation of spherical top spectra, JQSRT 59 (1998) 471--480.
\newblock \href {http://dx.doi.org/10.1016/S0022-4073(97)00106-4}
  {\path{doi:10.1016/S0022-4073(97)00106-4}}.

\bibitem{Wenger_2008a}
C.~Wenger, J.~P. Champion, V.~Boudon, The partition sum of methane at high
  temperature, JQSRT 109 (2008) 2697--2706.
\newblock \href {http://dx.doi.org/10.1016/j.jqsrt.2008.06.006}
  {\path{doi:10.1016/j.jqsrt.2008.06.006}}.

\bibitem{Rothman_2009}
L.~S. Rothman, I.~E. Gordon, A.~Barbe, D.~C. Benner, P.~E. Bernath, M.~Birk,
  V.~Boudon, L.~R. Brown, A.~Campargue, J.~P. Champion, K.~Chance, L.~H.
  Coudert, V.~Dana, V.~M. Devi, S.~Fally, J.~M. Flaud, R.~R. Gamache,
  A.~Goldman, D.~Jacquemart, I.~Kleiner, N.~Lacome, W.~J. Lafferty, J.~Y.
  Mandin, S.~T. Massie, S.~N. Mikhailenko, C.~E. Miller, N.~Moazzen-Ahmadi,
  O.~V. Naumenko, A.~V. Nikitin, J.~Orphal, V.~I. Perevalov, A.~Perrin,
  A.~Predoi-Cross, C.~P. Rinsland, M.~Rotger, M.~Simeckova, M.~A.~H. Smith,
  K.~Sung, S.~A. Tashkun, J.~Tennyson, R.~A. Toth, A.~C. Vandaele,
  J.~Vander~Auwera, The {HITRAN} 2008 molecular spectroscopic database, JQSRT
  110 (2009) 533--572.
\newblock \href {http://dx.doi.org/10.1016/j.jqsrt.2009.02.013}
  {\path{doi:10.1016/j.jqsrt.2009.02.013}}.

\bibitem{Jacquinet-Husson_2008}
N.~Jacquinet-Husson, N.~A. Scott, A.~Chedin, L.~Crepeau, R.~Armante,
  V.~Capelle, J.~Orphal, A.~Coustenis, C.~Boonne, N.~Poulet-Crovisier,
  A.~Barbee, M.~Birk, L.~R. Brown, C.~Camy-Peyret, C.~Claveau, K.~Chance,
  N.~Christidis, C.~Clerbaux, P.~F. Coheur, V.~Dana, L.~Daumont, M.~R.
  De~Backer-Barilly, G.~Di~Lonardo, J.~M. Flaud, A.~Goldman, A.~Hamdouni,
  M.~Hess, M.~D. Hurley, D.~Jacquemart, I.~Kleiner, P.~Kopke, J.~Y. Mandin,
  S.~Massie, S.~Mikhailenko, V.~Nemtchinov, A.~Nikitin, D.~Newnham, A.~Perrin,
  V.~I. Perevalov, S.~Pinnock, L.~Regalia-Jarlot, C.~P. Rinsland, A.~Rublev,
  F.~Schreier, L.~Schult, K.~M. Smith, S.~A. Tashkun, J.~L. Teffo, R.~A. Toth,
  V.~G. Tyuterev, J.~V. Auwera, P.~Varanasi, G.~Wagner, The geisa spectroscopic
  database: Current and future archive for earth and planetary atmosphere
  studies, JQSRT 109 (2008) 1043--1059.
\newblock \href {http://dx.doi.org/10.1016/j.jqsrt.2007.12.015}
  {\path{doi:10.1016/j.jqsrt.2007.12.015}}.

\bibitem{Nikitin_2005}
A.~Nikitin, J.~P. Champion, New ground state constants of
  $^{12}${CH}$_{3}$$^{35}${C}l and $^{12}${CH}$_{3}$$^{37}${C}l from global
  polyad analysis, J Mol Spectrosc 230 (2005) 168--173.
\newblock \href {http://dx.doi.org/10.1016/j.jms.2004.10.012}
  {\path{doi:10.1016/j.jms.2004.10.012}}.

\bibitem{Wenger_2000}
C.~Wenger, V.~Boudon, J.~P. Champion, G.~Pierre, Highly-spherical top data
  system ({HTDS}) software for spectrum simulation of octahedral {XY}$_6$
  molecules, JQSRT 66 (2000) 1--16.
\newblock \href {http://dx.doi.org/10.1016/S0022-4073(99)00161-2}
  {\path{doi:10.1016/S0022-4073(99)00161-2}}.

\bibitem{Griffith_1962}
J.~S. Griffith, The irreducible tensor method for molecular symmetry groups,
  Prentice-Hall, Englewood Cliffs, New York, NY, 1962.

\bibitem{Nikitin_2009a}
A.~V. Nikitin, J.~P. Champion, R.~A.~H. Butler, L.~R. Brown, I.~Kleiner, Global
  modeling of the lower three polyads of {PH}$_3$: Preliminary results, J Mol
  Spectrosc 256 (2009) 4--16.
\newblock \href {http://dx.doi.org/10.1016/j.jms.2009.01.008}
  {\path{doi:10.1016/j.jms.2009.01.008}}.

\bibitem{Nikitin_2008}
A.~V. Nikitin, Vibrational energy levels of methyl chloride calculated from
  full dimensional ab initio potential energy surface, J Mol Spectrosc 252
  (2008) 17--21.
\newblock \href {http://dx.doi.org/10.1016/j.jms.2008.06.001}
  {\path{doi:10.1016/j.jms.2008.06.001}}.

\bibitem{Nikitin_2011a}
A.~V. Nikitin, L.~Daumont, X.~Thomas, L.~Regalia, M.~Rey, V.~G. Tyuterev, L.~R.
  Brown, Preliminary assignments of $2\nu_3-\nu_4$ hot band of
  $^{12}${CH}$_{4}$ in the $2\mu$m transparency window from long-path {FTS}
  spectra, J Mol Spectrosc 268 (2011) 93--106.
\newblock \href {http://dx.doi.org/10.1016/j.jms.2011.04.002}
  {\path{doi:10.1016/j.jms.2011.04.002}}.

\bibitem{Nikitin_2011b}
A.~V. Nikitin, An efficient code for calculation of the 6{C}, 9{C} and 12{C}
  symbols for {C}$_{3v}$, {T}$_d$ and {O}$_h$ point groups, Computer Physics
  Communication in press (2011) CPC--D--11--00282.

\bibitem{WATSON_1967}
J.~K.~G. Watson, Determination of centrifugal distortion coefficients of
  asymmetric-top molecules, J Chem Phys 46 (1967) 1935--.
\newblock \href {http://dx.doi.org/10.1063/1.1840957}
  {\path{doi:10.1063/1.1840957}}.

\bibitem{KLEIN_1974}
D.~J. Klein, Degenerate perturbation theory, J Chem Phys 61 (1974) 786--798.
\newblock \href {http://dx.doi.org/10.1063/1.1682018}
  {\path{doi:10.1063/1.1682018}}.

\bibitem{Tyuterev_1975}
V.~G. Tyuterev, Effective hamiltonians, in: Intramolecular interactions and
  Infrared spectra of Atmospheric gases, Acad. Sci., Tomsk, 1975, pp. 3--46.

\bibitem{TYUTEREV_1980}
V.~G. Tyuterev, V.~I. Perevalov, Generalized contact transformations of a
  hamiltonian with a quasi-degenerate zero-order approximation - application to
  accidental vibration-rotation resonances in molecules, Chem Phys Lett 74
  (1980) 494--502.
\newblock \href {http://dx.doi.org/10.1016/0009-2614(80)85260-2}
  {\path{doi:10.1016/0009-2614(80)85260-2}}.

\bibitem{Makushkin_1984}
Y.~S. Makushkin, V.~G. Tyuterev, Perturbation Methods and Effective
  Hamiltonians in Molecular Spectroscopy, Nauka, Novosibirsk [in Russian],
  1984.

\bibitem{JORGENSEN_1975}
F.~Jorgensen, Effective hamiltonians, Mol Phys 29 (1975) 1137--1164.
\newblock \href {http://dx.doi.org/10.1080/00268977500100971}
  {\path{doi:10.1080/00268977500100971}}.

\bibitem{PEREVALOV_1982b}
V.~I. Perevalov, V.~G. Tyuterev, Reduction of the centrifugal-distortion
  hamiltonian of asymmetric-top molecules in the case of accidental resonances
  - 2 interacting states - lower-order terms, J Mol Spectrosc 96 (1982) 56--76.
\newblock \href {http://dx.doi.org/10.1016/0022-2852(82)90214-4}
  {\path{doi:10.1016/0022-2852(82)90214-4}}.

\bibitem{PEREVALOV_1984a}
V.~I. Perevalov, V.~G. Tyuterev, B.~I. Zhilinskii, Ambiguity of spectroscopic
  parameters in the case of accidental vibration-rotation resonances in
  tetrahedral molecules - {R}$^2${J} and {R}$^2${J}$^2$ terms for {E}-{F}$_2$
  interacting states, Chem Phys Lett 104 (1984) 455--461.
\newblock \href {http://dx.doi.org/10.1016/0009-2614(84)85622-5}
  {\path{doi:10.1016/0009-2614(84)85622-5}}.

\bibitem{TYUTEREV_1986}
V.~G. Tyuterev, J.~P. Champion, G.~Pierre, V.~I. Perevalov, Parameters of
  reduced hamiltonian and invariant parameters of interacting {E} and {F}$_2$
  fundamentals of tetrahedral molecules - $\nu_2$ and $\nu_4$ bands of
  $^{12}${CH}$_{4}$ and $^{28}${SiH}$_{4}$, J Mol Spectrosc 120 (1986) 49--78.
\newblock \href {http://dx.doi.org/10.1016/0022-2852(86)90070-6}
  {\path{doi:10.1016/0022-2852(86)90070-6}}.

\bibitem{LOBODENKO_1987a}
E.~I. Lobodenko, O.~N. Sulakshina, V.~I. Perevalov, V.~G. Tyuterev, Reduced
  effective hamiltonian for coriolis-interacting $\nu_n$ and $\nu_t$
  fundamentals of {C}$_{3v}$ molecules, J Mol Spectrosc 126 (1987) 159--170.
\newblock \href {http://dx.doi.org/10.1016/0022-2852(87)90086-5}
  {\path{doi:10.1016/0022-2852(87)90086-5}}.

\bibitem{Aliev_1985}
M.~Aliev, J.~Watson, Higher-order effects in the vibration-rotation spectra of
  semirigid molecules, in: K.~Rao (Ed.), Molecular Spectroscopy: Modern
  Research, Vol. III, Academic Press, New York, 1985, pp. 1--67.

\bibitem{PEREVALOV_1984}
V.~I. Perevalov, V.~G. Tyuterev, B.~I. Zhilinskii, Reduced
  effective-hamiltonians for degenerate vibrational-states of methane-type
  molecules, J Mol Spectrosc 103 (1984) 147--159.
\newblock \href {http://dx.doi.org/10.1016/0022-2852(84)90153-X}
  {\path{doi:10.1016/0022-2852(84)90153-X}}.

\bibitem{TYUTEREV_1990}
V.~G. Tyuterev, J.~P. Champion, G.~Pierre, Reduced effective-hamiltonians for
  degenerate excited vibrational-states of tetrahedral molecules - application
  to $2\nu_2$, $\nu_2+\nu_4$ and $2\nu_4$ of {CH}$_{4}$, Mol Phys 71 (1990)
  995--1020.
\newblock \href {http://dx.doi.org/10.1080/00268979000102281}
  {\path{doi:10.1080/00268979000102281}}.

\bibitem{Sarka_2000}
K.~Sarka, J.~Demaison, Perturbation theory, effective hamiltonians and force
  constants, in: P.~Jensen, P.~R. Bunker (Eds.), Computational Molecular
  Spectroscopy, Wiley, Chichester, 2000.

\bibitem{Tyuterev_2002}
V.~G. Tyuterev, Effective hamiltonians and perturbation theory for quantum
  bound states of nuclear motion in molecules, in: G.~Gaeta (Ed.), Symmetry and
  Perturbation Theory, World Scientic Publishing, New Jersey, 2002, p. 254.

\bibitem{Rey_2003}
M.~Rey, V.~Boudon, C.~Wenger, G.~Pierre, B.~Sartakov, Orientation of {O}(3) and
  {SU}(2) $\otimes$ {C}$_1$ representations in cubic point groups {O}$_h$,
  {T}$_d$ for application to molecular spectroscopy, J Mol Spectrosc 219 (2003)
  313--325.
\newblock \href {http://dx.doi.org/10.1016/S0022-2852(03)00056-0}
  {\path{doi:10.1016/S0022-2852(03)00056-0}}.

\bibitem{Tashkun_2010}
S.~A. Tashkun, V.~I. Perevalov, R.~V. Kochanov, A.~W. Liu, S.~M. Hu, Global
  fittings of $^{14}${N}$^{15}${N}$^{16}${O} and $^{15}${N}$^{14}${N}$^{16}${O}
  vibrational-rotational line positions using the effective hamiltonian
  approach, JQSRT 111 (2010) 1089--1105.
\newblock \href {http://dx.doi.org/10.1016/j.jqsrt.2010.01.010}
  {\path{doi:10.1016/j.jqsrt.2010.01.010}}.

\bibitem{Tashkun_1993}
S.~A. Tashkun, V.~G. Tyuterev, {GIP}: a program for experimental data reduction
  in molecular spectroscopy, Vol. 2205, Proc. SPIE, 1993, pp. 188--191.

\bibitem{Herman_2007}
M.~Herman, The acetylene ground state saga, Mol Phys 105 (2007) 2217--2241.
\newblock \href {http://dx.doi.org/10.1080/00268970701518103}
  {\path{doi:10.1080/00268970701518103}}.

\bibitem{Amyay_2011}
B.~Amyay, M.~Herman, A.~Fayt, A.~Campargue, S.~Kassi, Acetylene,
  $^{12}${C}$_2${H}$_2$: Refined analysis of {CRDS} spectra around 1.52 $\mu$m,
  J Mol Spectrosc 267 (2011) 80--91.
\newblock \href {http://dx.doi.org/10.1016/j.jms.2011.02.015}
  {\path{doi:10.1016/j.jms.2011.02.015}}.

\bibitem{JENSEN_1988d}
P.~Jensen, A variational calculation of the rotation vibration energies for
  {H}$_2${O} from abinitio data, J Mol Struct 190 (1988) 149--161.
\newblock \href {http://dx.doi.org/10.1016/0022-2860(88)80280-1}
  {\path{doi:10.1016/0022-2860(88)80280-1}}.

\bibitem{JENSEN_1988e}
P.~Jensen, Calculation of rotation vibration linestrengths for
  triatomic-molecules using a variational approach - application to the
  fundamental bands of {CH}$_2$, J Mol Spectrosc 132 (1988) 429--457.
\newblock \href {http://dx.doi.org/10.1016/0022-2852(88)90338-4}
  {\path{doi:10.1016/0022-2852(88)90338-4}}.

\bibitem{Schwenke_1996}
D.~W. Schwenke, Variational calculations of rovibrational energy levels and
  transition intensities for tetratomic molecules, J Phys Chem 100 (1996)
  2867--2884.
\newblock \href {http://dx.doi.org/10.1021/jp9525447}
  {\path{doi:10.1021/jp9525447}}.

\bibitem{Bowman_2010}
J.~M. Bowman, B.~J. Braams, S.~Carter, C.~Chen, G.~Czako, B.~Fu, X.~Huang,
  E.~Kamarchik, A.~R. Sharma, B.~C. Shepler, Y.~Wang, Z.~Xie, Ab-initio-based
  potential energy surfaces for complex molecules and molecular complexes, J
  Phys Chem Letters 1 (2010) 1866--1874.
\newblock \href {http://dx.doi.org/10.1021/jz100626h}
  {\path{doi:10.1021/jz100626h}}.

\bibitem{Yurchenko_2011}
S.~N. Yurchenko, R.~J. Barber, J.~Tennyson, A variationally computed line list
  for hot {NH}$_3$, Monthly Notices of the Royal Astronomical Society 413
  (2011) 1828--1834.
\newblock \href {http://dx.doi.org/10.1111/j.1365-2966.2011.18261.x}
  {\path{doi:10.1111/j.1365-2966.2011.18261.x}}.

\bibitem{CARTER_1986}
S.~Carter, N.~C. Handy, The variational method for the calculation of
  ro-vibrational energy-levels, Computer Physics Reports 5 (1986) 115--172.
\newblock \href {http://dx.doi.org/10.1016/0167-7977(86)90006-7}
  {\path{doi:10.1016/0167-7977(86)90006-7}}.

\bibitem{Nikitin_2011c}
A.~V. Nikitin, M.~Rey, V.~G. Tyuterev, Rotational and vibrational energy levels
  of methane calculated from a new potential energy surface, Chem Phys Lett 501
  (2011) 179--186.
\newblock \href {http://dx.doi.org/10.1016/j.cplett.2010.11.008}
  {\path{doi:10.1016/j.cplett.2010.11.008}}.

\bibitem{Zuniga_2007}
J.~Zuniga, J.~A.~G. Picon, A.~Bastida, A.~Requena, Optimal internal
  coordinates, vibrational spectrum, and effective hamiltonian for ozone, J
  Chem Phys 126 (2007) 244305.
\newblock \href {http://dx.doi.org/10.1063/1.2743441}
  {\path{doi:10.1063/1.2743441}}.

\bibitem{Schwenke_2002}
D.~W. Schwenke, Towards accurate ab initio predictions of the vibrational
  spectrum of methane, Spectrochim Acta Part A 58 (2002) 849--861.
\newblock \href {http://dx.doi.org/10.1016/S1386-1425(01)00673-4}
  {\path{doi:10.1016/S1386-1425(01)00673-4}}.

\bibitem{Matyus_2009}
E.~Matyus, G.~Czako, A.~G. Csaszar, Toward black-box-type full- and
  reduced-dimensional variational (ro)vibrational computations, J Chem Phys 130
  (2009) 134112.
\newblock \href {http://dx.doi.org/10.1063/1.3076742}
  {\path{doi:10.1063/1.3076742}}.

\bibitem{Tyuterev_2004}
V.~G. Tyuterev, L.~Regalia-Jarlot, D.~W. Schwenke, S.~A. Tashkun, Y.~G. Borkov,
  Global variational calculations of high-resolution rovibrational spectra:
  isotopic effects, intensity anomalies and experimental confirmations for
  {H}$_2${S}, {HDS}, {D}$_2${S} molecules, Comptes Rendus Physique 5 (2004)
  189--199.
\newblock \href {http://dx.doi.org/10.1016/j.crhy.2004.01.017}
  {\path{doi:10.1016/j.crhy.2004.01.017}}.

\bibitem{Light_2000}
J.~C. Light, T.~Carrington, Discrete-variable representations and their
  utilization, Adv Chem Phys 114 (2000) 263--310.
\newblock \href {http://dx.doi.org/10.1002/9780470141731.ch4}
  {\path{doi:10.1002/9780470141731.ch4}}.

\bibitem{Wang_2011b}
Y.~M. Wang, J.~M. Bowman, Ab initio potential and dipole moment surfaces for
  water. ii. local-monomer calculations of the infrared spectra of water
  clusters, J Chem Phys 134 (2011) 154510.
\newblock \href {http://dx.doi.org/10.1063/1.3579995}
  {\path{doi:10.1063/1.3579995}}.

\bibitem{Wang_2011c}
Y.~M. Wang, X.~C. Huang, B.~C. Shepler, B.~J. Braams, J.~M. Bowman, Flexible,
  ab initio potential, and dipole moment surfaces for water. i. tests and
  applications for clusters up to the 22-mer, J Chem Phys 134 (2011) 094509.
\newblock \href {http://dx.doi.org/10.1063/1.3554905}
  {\path{doi:10.1063/1.3554905}}.

\bibitem{Tennyson_2000}
J.~Tennyson, Variational calculations of vibration-rotation spectra, in:
  P.~Jensen, P.~R. Bunker (Eds.), Computational Molecular Spectroscopy, Wiley,
  Chichester, 2000, pp. 305--324.

\bibitem{Tennyson_1992}
J.~Tennyson, S.~Miller, J.~R. Henderson, Methods in Computational Chemistry,
  Vol.~4, Plenum, New York, 2000.

\bibitem{Wang_2002}
X.~G. Wang, T.~Carrington, New ideas for using contracted basis functions with
  a lanczos eigensolver for computing vibrational spectra of molecules with
  four or more atoms, J Chem Phys 117 (2002) 6923--6934.
\newblock \href {http://dx.doi.org/10.1063/1.1506911}
  {\path{doi:10.1063/1.1506911}}.

\bibitem{Mandelshtam_1997}
V.~A. Mandelshtam, H.~S. Taylor, The quantum resonance spectrum of the
  {H}$_3$$^+$ molecular ion for $j=0$. an accurate calculation using filter
  diagonalization, J Chem Soc Faraday T 93 (1997) 847--860.
\newblock \href {http://dx.doi.org/10.1039/a607010h}
  {\path{doi:10.1039/a607010h}}.

\bibitem{Tyuterev_2003a}
V.~G. Tyuterev, Recent advances in global variational and effective
  calculations of the line positions and intensities for triatomic molecules:
  some features of a new generation of spectroscopic databanks., Atmos Ocean
  Optics 16 (2003) 220--230.

\bibitem{VanVleck_1929}
J.~H. Van~Vleck, On sigma-type doubling and electron spin in the spectra of
  diatomic molecules, Phys Rev 33 (1929) 0467--0506.
\newblock \href {http://dx.doi.org/10.1103/PhysRev.33.467}
  {\path{doi:10.1103/PhysRev.33.467}}.

\bibitem{SIBERT_1988a}
E.~L. Sibert, Theoretical studies of vibrationally excited polyatomic molecules
  using canonical vanvleck perturbation theory, J Chem Phys 88 (1988)
  4378--4390.
\newblock \href {http://dx.doi.org/10.1063/1.453797}
  {\path{doi:10.1063/1.453797}}.

\bibitem{Wang_1999}
X.~G. Wang, E.~L. Sibert, A nine-dimensional perturbative treatment of the
  vibrations of methane and its isotopomers, J Chem Phys 111 (1999) 4510--4522.
\newblock \href {http://dx.doi.org/10.1063/1.480271}
  {\path{doi:10.1063/1.480271}}.

\bibitem{Dubernet_2010}
M.~L. Dubernet, V.~Boudon, J.~L. Culhane, M.~S. Dimitrijevic, A.~Z. Fazliev,
  C.~Joblin, F.~Kupka, G.~Leto, P.~Le~Sidaner, P.~A. Loboda, H.~E. Mason, N.~J.
  Mason, C.~Mendoza, G.~Mulas, T.~J. Millar, L.~A. Nunez, V.~I. Perevalov,
  N.~Piskunov, Y.~Ralchenko, G.~Rixon, L.~S. Rothman, E.~Roueff, T.~A.
  Ryabchikova, A.~Ryabtsev, S.~Sahal-Brechot, B.~Schmitt, S.~Schlemmer,
  J.~Tennyson, V.~G. Tyuterev, N.~A. Walton, V.~Wakelam, C.~J. Zeippen, Virtual
  atomic and molecular data centre, JQSRT 111 (2010) 2151--2159.
\newblock \href {http://dx.doi.org/10.1016/j.jqsa.2010.05.004}
  {\path{doi:10.1016/j.jqsa.2010.05.004}}.

\end{thebibliography}

\end{document}